\documentclass[epjCONF,columns]{svjour} %for 2 columns format
\usepackage{graphics}
\usepackage[varg]{txfonts} % Times fonts
\usepackage[latin1]{inputenc}
\usepackage{epsfig}
\usepackage{bm}
\session-title{Hadron Collider Physics symposium 2011}
\def\Journal#1#2#3#4{{#1} {\bf #2}, #3 (#4)}

\def\PLB{{\em Phys. Lett.}  B}

\def\PRD{{\em Phys. Rev.} D}

\def\be{\begin{equation}}
\def\ee{\end{equation}}
\def\bea{\begin{eqnarray}}
\def\eea{\end{eqnarray}}

\def\dphiLL{\mathrm{\Delta\phi_{\ell\ell}}}

\def\dRLL{\mathrm{\Delta R_{\ell\ell}}}
\def\mH{\mathrm{m_{\rm H}}}
\def\etmiss{\mathrm{E_{T}^{miss}}}
\def\pT{\mathrm{p_{T}}}
\def\sigmaRatio{$\sigma_{\mathrm{95}}/\sigma_{\mathrm{SM}}$}

\begin{document}
\title{High Mass Standard Model Higgs searches at the Tevatron}
\author{Konstantinos A. Petridis\inst{1}\fnmsep\thanks{\email{kap01@fnal.gov}} 
on behalf of the CDF and D0 collaborations}
\institute{The University of Manchester}
\abstract{ We present the results of searches for the Standard Model
  Higgs boson decaying predominantly to $\mathrm{W^+W^-}$ pairs, at a
  center-of-mass energy of $\sqrt{s}=$1.96\,TeV, using up to
  8.2\,fb$^{-1}$ of data collected with the CDF and D0 detectors at
  the Fermilab Tevatron collider. The analysis techniques and the
  various channels considered are discussed. These searches result in
  exclusions across the Higgs mass range of 156.5$<\mH<$173.7\,GeV for CDF
  and 161$<\mH<$170\,GeV for D0. }
\maketitle

\section{Introduction}
\label{sec:intro}
The search for the mechanism of electroweak symmetry breaking and a
Standard Model (SM) Higgs boson has been a major goal of particle
physics for many years. Within the Higgs sector of the SM, the mass of
the Higgs boson ($\mH$) is a free parameter. Constraints on $\mH$ come
from direct searches at the LEP experiments\,\cite{ref:LEPexper} which
conclude that $\mH$$>$114.4\,GeV at 95\% Confidence Level (CL) and
more recently from combined searches of the ATLAS and CMS
experiments using up to 2.3\,fb$^{-1}$ data yielding an exclusion
of 141$<\mH<$476\,GeV at 95\% CL\,\cite{ref:Gigi}. Additionally,
indirect constraints using precision electroweak measurements 
require $\mH$$<$185\,GeV at 95\% CL\,\cite{ref:LEPEWK}.

Higgs searches at the Tevatron collider are a subject of intense
study. By combining CDF and D0 results, the SM Higgs mass range
between 158 and 175\,GeV was excluded at 95\% CL using up to
8.1\,fb$^{-1}$ of data\,\cite{ref:TeVWinter2011}.  

These proceedings present the status of the searches by CDF and D0 for
a SM Higgs boson decaying predominantly to a pair of W bosons and using
up to 8.2\,fb$^{-1}$ of data collected until the summer of 2011. All limits
will be given at 95\% CL

\section{Higgs boson production and decays}
\label{sec:higgsprod}
At the Tevatron, the dominant production mode is via the gluon fusion
process, $\mathrm{gg\to H}$, with a cross section ranging between
840-190\,fb for a Higgs mass between 130-200\,GeV. The associated
production, $\mathrm{q\bar{q}\to VH}$ ($\mathrm{V=W,Z}$), and vector
boson fusion, $\mathrm{q\bar{q}\to Hq\bar{q}}$, processes with cross
sections ranging between 180-32\,fb and 57-22\,fb respectively for the
aforementioned Higgs mass range, are also considered in order to
maximise the sensitivity of these searches.

For a Higgs mass $\mH$$>$135\,GeV, the main decay mode is to a pair of
$\rm W$ bosons while for $\mH$$<$135\,GeV Higgs decays mainly to a
pair of b-quarks. This distinction is what defines the high mass and
low mass Higgs searches at the Tevatron.

\section{Search channels}
\label{sec:channels}
The high mass searches at CDF and D0 require at least one electron or
a muon in the final state in order to suppress the QCD
background. Given this requirement all possible decay modes are
considered to maximise the signal acceptance. The di-lepton+missing
transvere energy ($\etmiss$) channel requires two electrons or muons
(plus neutrinos) of opposite charge in the final state. This
represents a small $\mathrm{WW}$ decay branching ratio, $\approx6\%$
(including $\tau\to e,\mu$ decays), but a clean signature offering the
highest sensitivity of all the high mass channels. Decays of one of
the $\rm W$ bosons to a $\tau$ lepton and the other to an electron or
muon, with a subsequent hadronic decay of the $\tau$ ($\tau_{\rm h})$
are also considered offering an additional branching ratio of
$\approx4\%$. The lepton+jets channel requires one $\rm W$ boson to
decay hadronically and the other leptonically. This represents a
signficant $\mathrm{WW}$ decay branching ratio, $\approx30\%$, however
suffers from a large $\rm W$+jets background. Dedicated
$\mathrm{q\bar{q}\to VH}$ searches are also performed by looking for
the SM suppressed signature of at least two leptons (electrons, muons)
with one same charge pair originating from leptonic decays of the
three vector bosons in the final state.

\section{Search strategy}
\label{sec:strategy}
The high mass analyses are each split into categories based on the
reconstructed lepton flavour, quality or jet multiplicity. This is
done in order to take advantage of the differences in the detector
response between the various lepton types, and between the kinematics
of the signal production mechanisms and the background processes.

\begin{figure*}
  \begin{center}
    \includegraphics[scale=0.38]{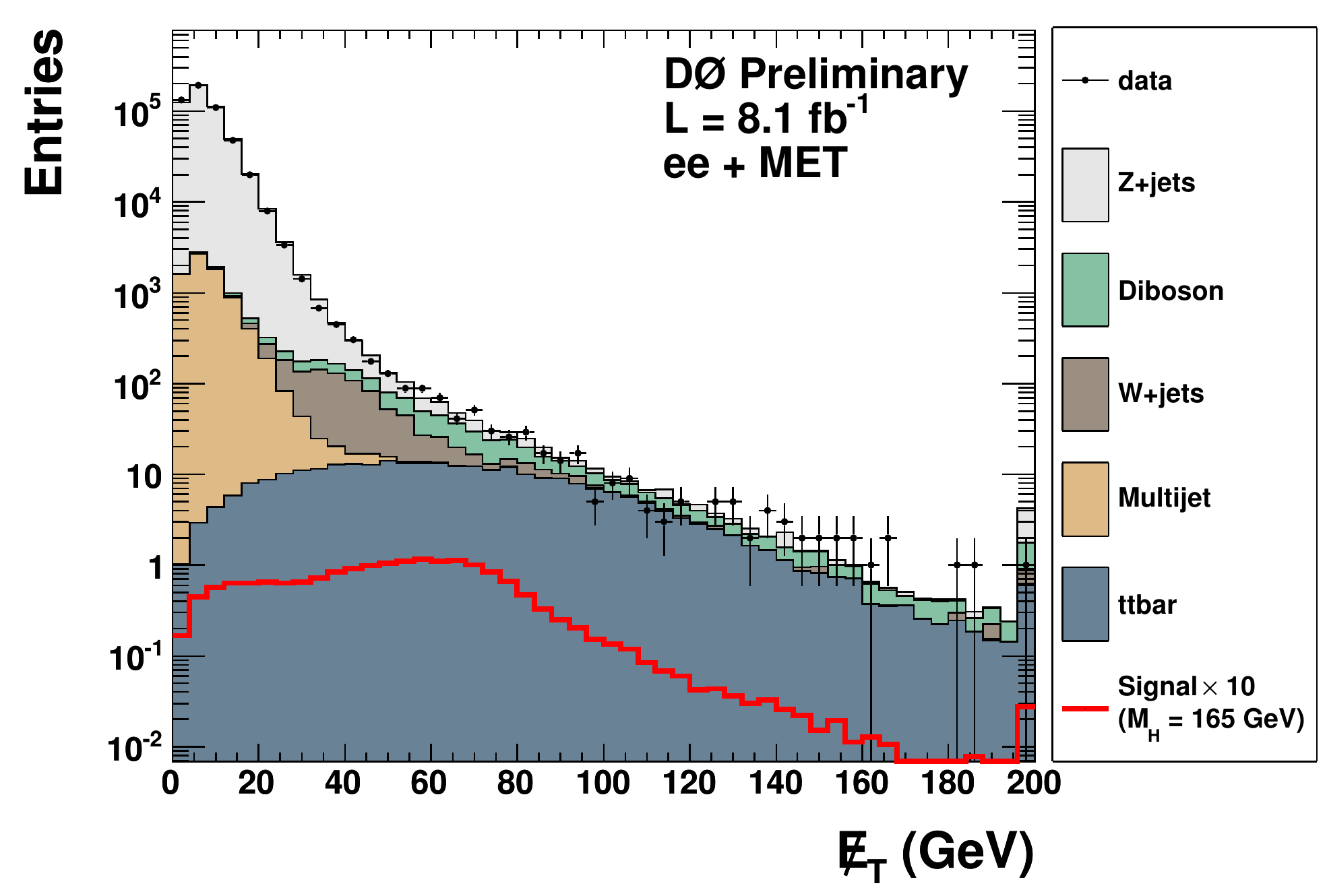}
    \includegraphics[scale=0.38]{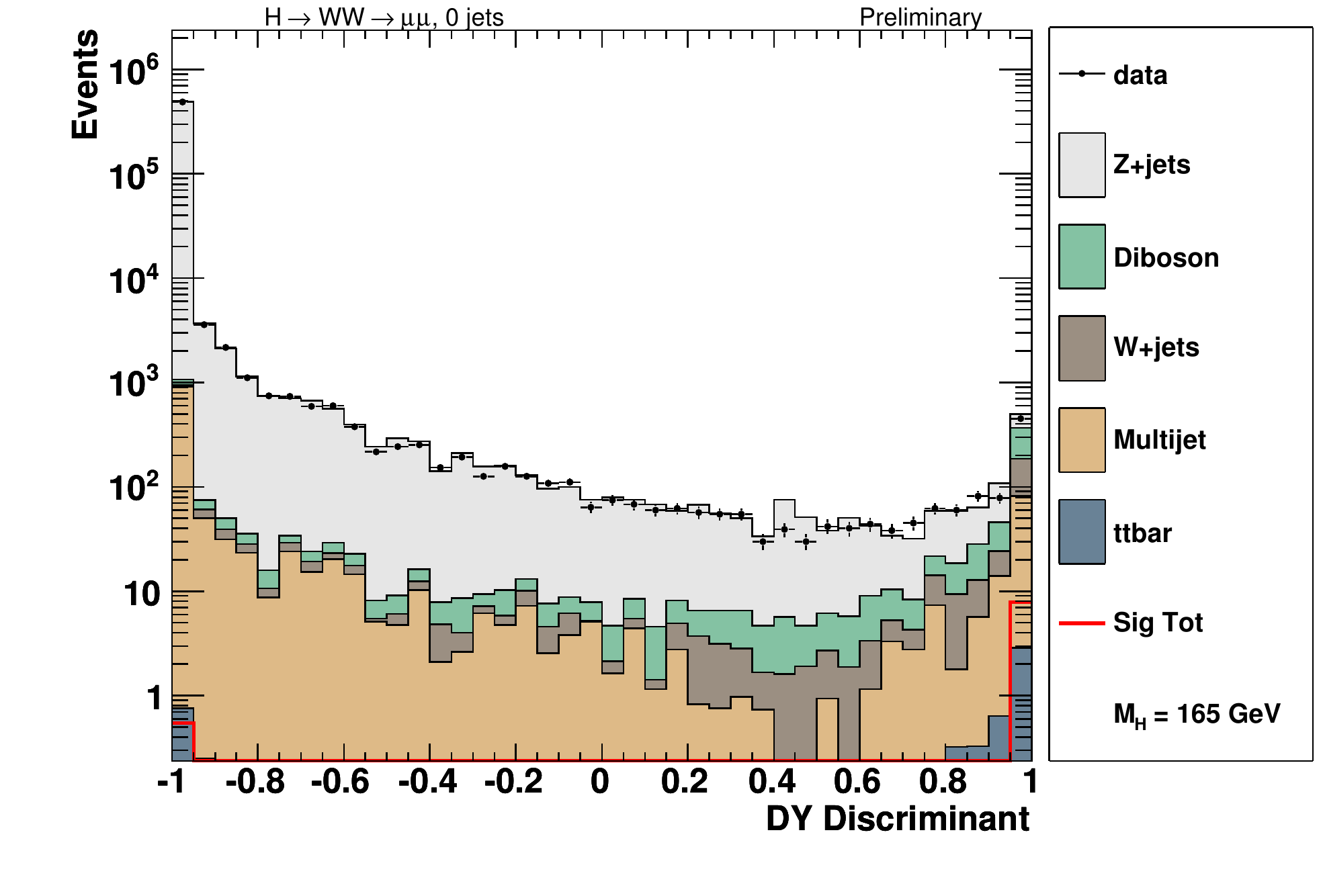}
    \put(-115, +130){ \bf \small {D\O\ Preliminary} }
    \put(-115, +120){ \bf \small L=8.1\,fb$^{-1}$ }
    \put(-115, +110){ \bf \small {$\bm{\mu\mu}$+MET}}
    \caption{Left: Distribution of $\etmiss$ di-electron+$\etmiss$
      events. Right: MVA discriminant trained to discriminate between
      signal and $\rm Z$+jets in di-muon+$\etmiss$ events.
      \label{fig:dy_bdt}}
  \end{center}
\end{figure*}

As the signal final states contain neutrinos, selections based on
large $\etmiss$ are used which account for energy mis-measurements in
the detectors. D0 employs Multivariate (MVA) techniques at this stage
to reduce low $\etmiss$ backgrounds while maximising signal
acceptance, by taking advantage of the correlations between the input
variables. Selecting events with a high MVA response score, improves
the sensitivity by up to 30\% compared to conventional square
selections.  Figure\,\ref{fig:dy_bdt} shows the $\etmiss$ distribution
for di-electron+$\etmiss$ events and the response of the MVA trained
against the large $\rm Z$+jets background.

Even after all selections, the signal/background (S/B) ratio is still
in the range of 0.1-1\% for $\mH$=165\,GeV. This requires the use of MVA
discriminants, either boosted decision trees or artificial neural
networks, to provide further discriminating power by using their
response to derive limits on the Higgs boson yield for the cases where
no signal-like excess is observed.

\subsection{Di-lepton+$\etmiss$ channel}\label{sec:osDilepton}
\begin{figure*}
  \begin{center}
    \includegraphics[scale=0.38]{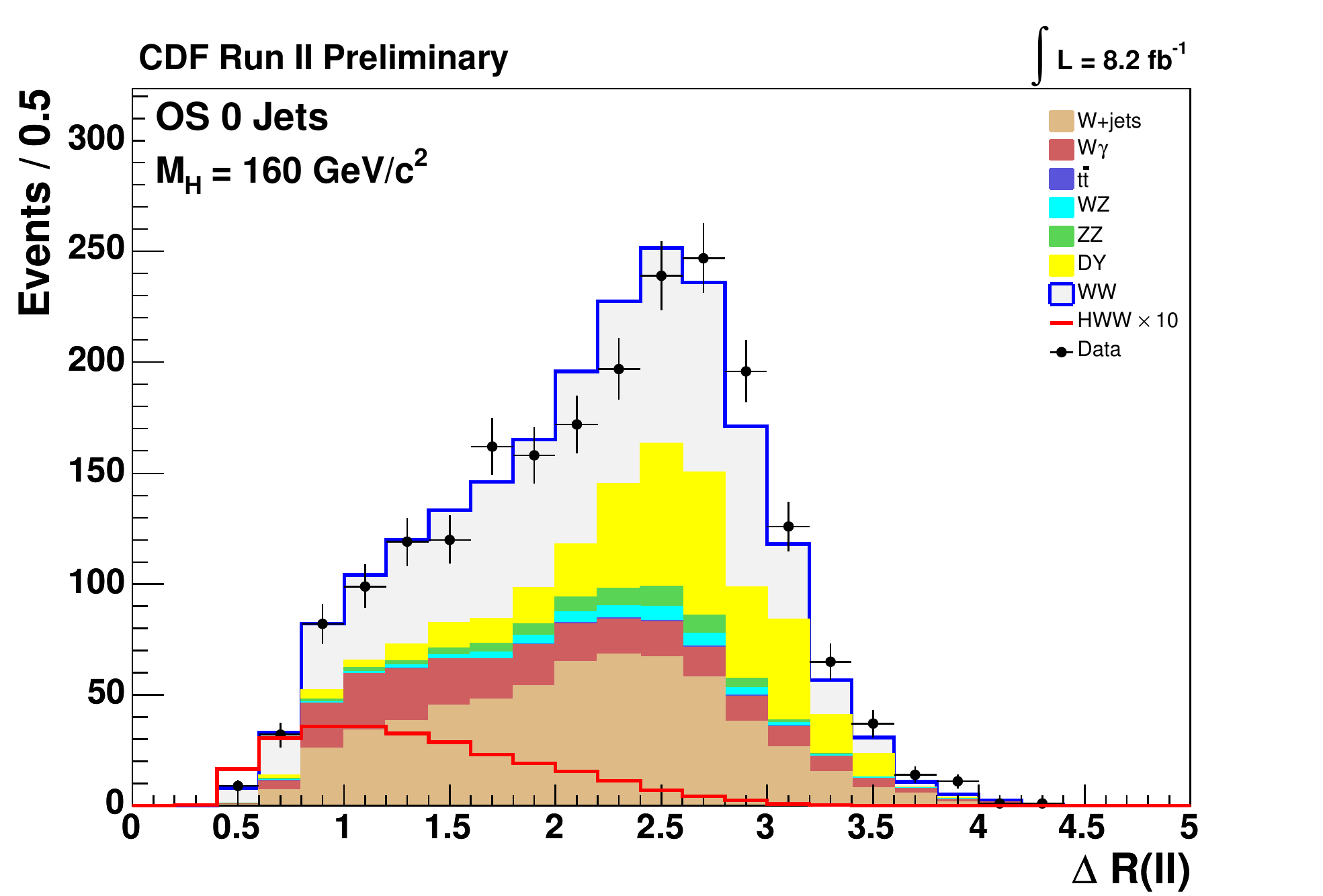}
    \includegraphics[scale=0.38]{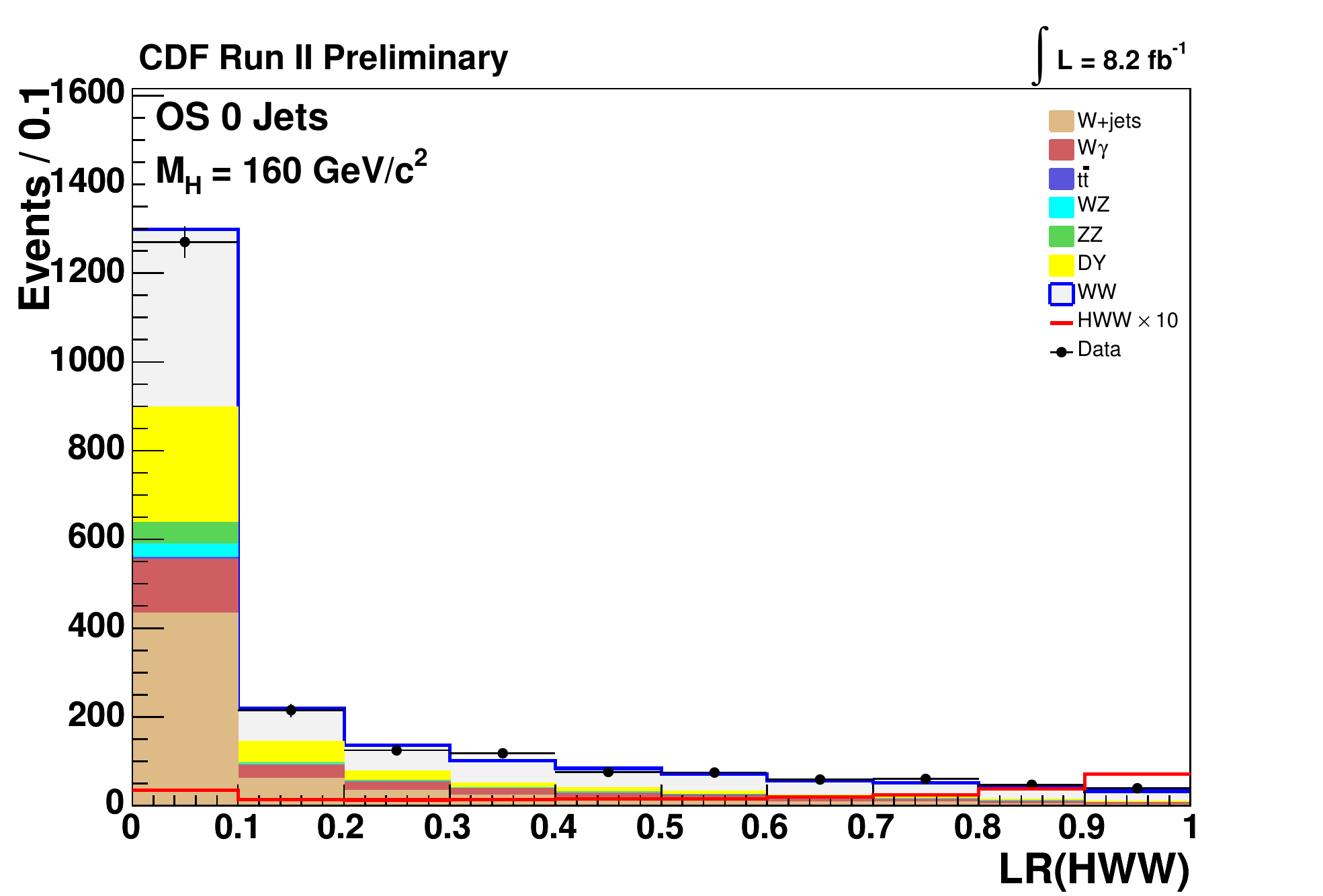}
    \caption{Left: $\dRLL$ distribution in di-lepton+$\etmiss$ events.
      Right: Distribution of the matrix element likelihood ratio in
      di-lepton+$\etmiss$ events. \label{fig:dRll}}
  \end{center}
\end{figure*}

The signature of this channel is two high p$_{\rm T}$, oppositely
charged, isolated leptons and $\etmiss$. A small angular separation is
also expected due to the spin-0 nature of the Higgs boson, which in
turn gives rise to the spin-correlation between the final-state
leptons. In contrast, leptons from the irreducible $\mathrm{Z\to WW}$
background are predominantly back-to-back. Figure\,\ref{fig:dRll}
shows the $\mathrm{\Delta R_{\ell\ell}=\sqrt{\Delta\phi_{\ell\ell}^2+\Delta\eta_{\ell\ell}^2}}$
distribution for a di-lepton+$\etmiss$ data sample with zero
reconstructed jets. Full use of this topological distinction is made
by using variables such as $\dRLL$ and $\dphiLL$ as inputs to the MVA
discriminants. CDF also makes use of likelihood ratios constructed
from matrix-element probabilities as input variables to the MVA
discriminants. Such a likelihood ratio is shown on the right plot
of Figure\,\ref{fig:dRll}.

\begin{figure*}
  \begin{center}
    \includegraphics[scale=0.38]{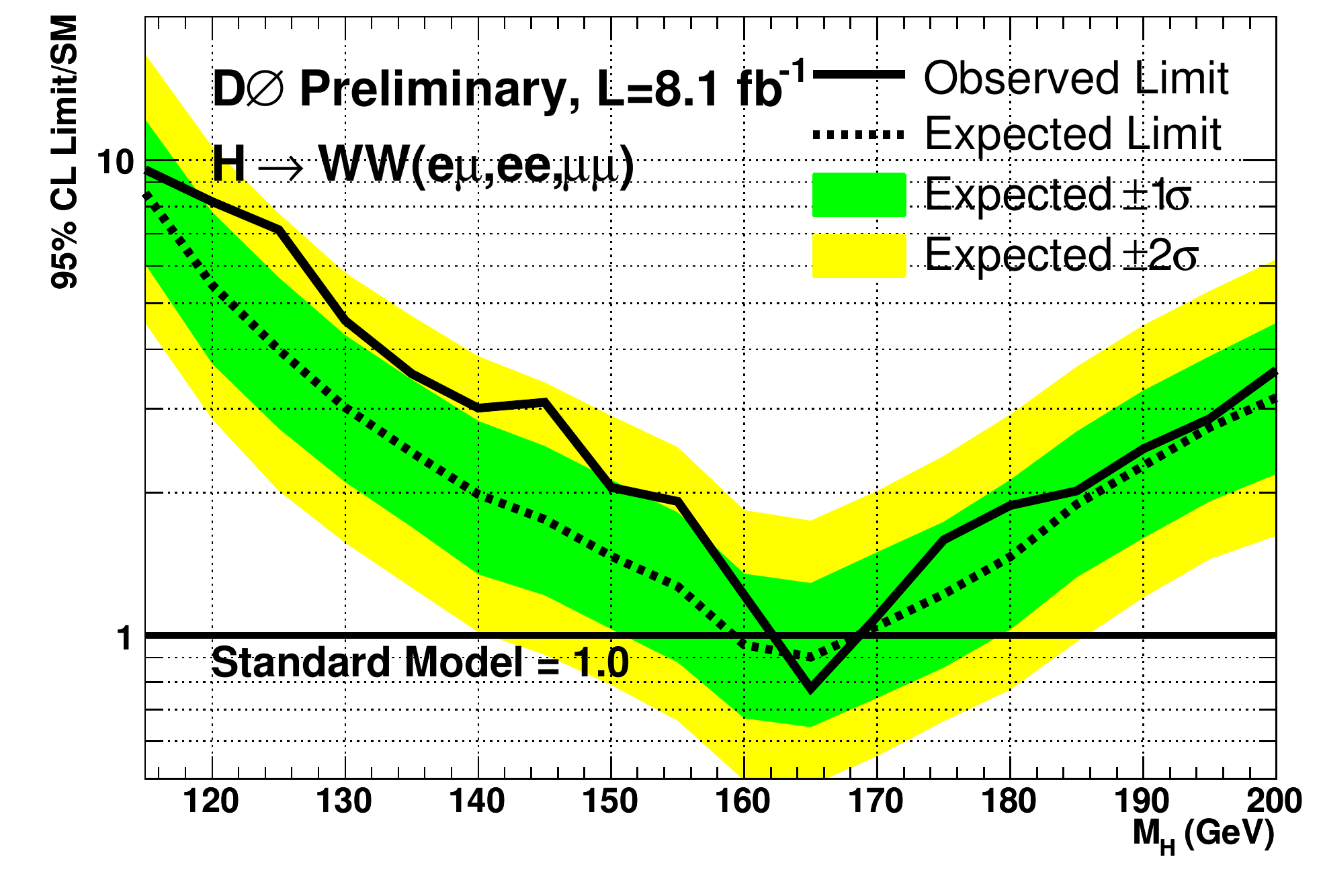}
    \includegraphics[scale=0.38]{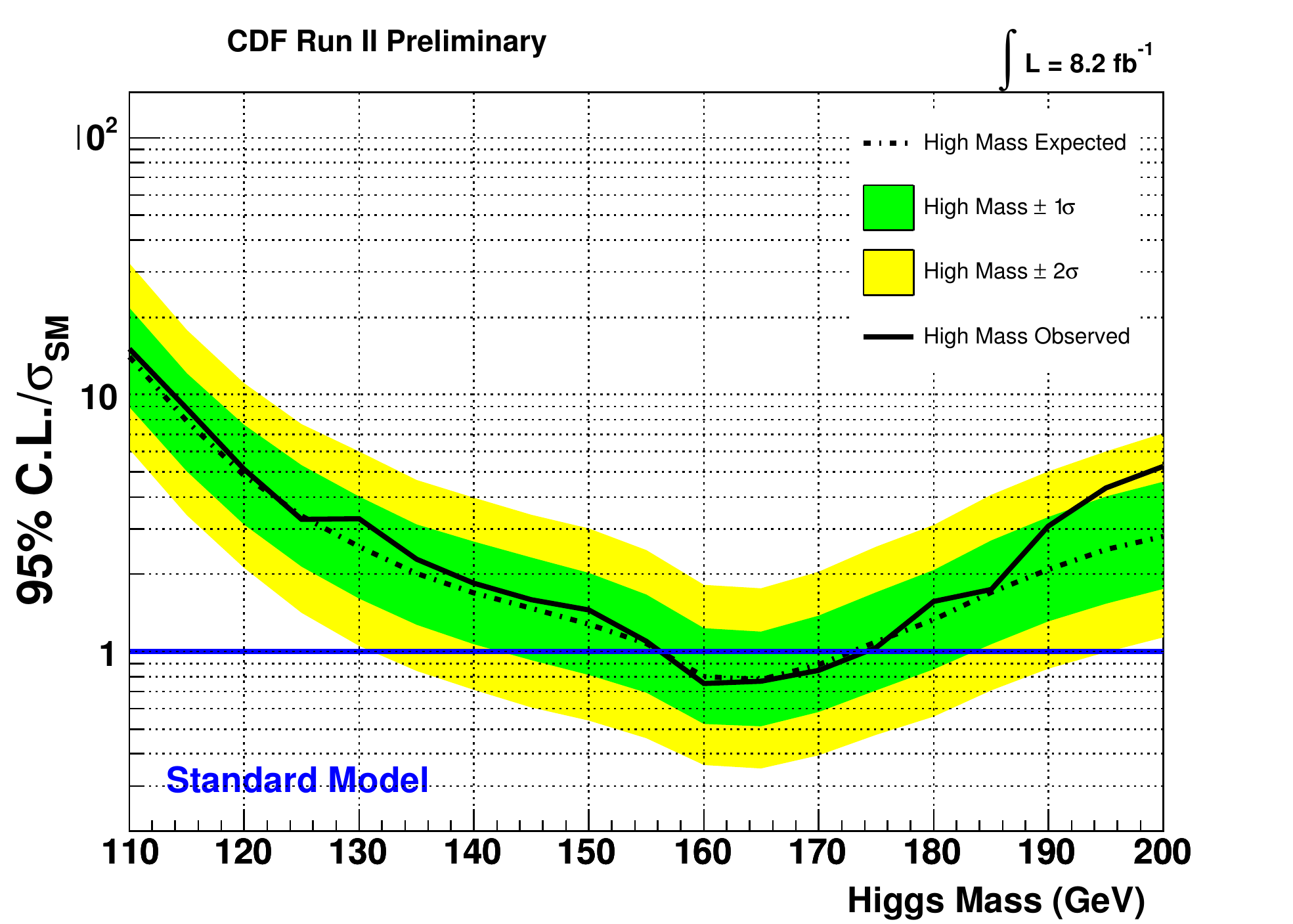}
    \caption{Combined limits using the Di-lepton+$\etmiss$ channels
      for D0 (left) and CDF (right). The CDF limit includes the same sign
      di-lepton and tri-lepton searches and the di-lepton+$\etmiss$
      with a $\tau_{\rm h}$ as described in sections
      \ref{sec:tauDilepton} and \ref{sec:ssDilepton}. \label{fig:hww_limits}}
  \end{center}
\end{figure*}

This channel is further split according to jet multiplicity into 0, 1
or more than 1 reconstructed jets in the final state. This enables the
training of the MVA discriminant to focus on the different signal and
background compositions in each of the jet multiplicity categories,
such as $\mathrm{WW}$ for the 0 jet, $\mathrm{Z+jets}$ for the 1 jet
and $\mathrm{t\bar{t}}$ for the 2 jet where b-tagging information is
used to suppress this background.  The statistical analysis of the MVA
final discriminant does not exhibit any excess with respect to the
background expectations and limits are set as shown in
Figure\,\ref{fig:hww_limits}. It should be noted that the CDF limit
was obtained including the same sign di-lepton and tri-lepton searches
described in sections \ref{sec:tauDilepton} and \ref{sec:ssDilepton}. 

\subsection{Di-lepton+$\etmiss$ channel with a $\tau_{\rm h}$}\label{sec:tauDilepton}
Additional signal acceptance can be obtained by considering
hadronically decaying $\tau$ leptons from W
decays. Muon (electron)+$\tau_{\rm h}$ channels are a new addition to
the D0 high mass program. In contrast to CDF, D0 splits these channels
into $\leq$1 or $>$1 jets due to the different signal production
mechanisms and background compositions. The $\leq$1 jet category does
not include an electron+$\tau_{\rm h}$ final state due to the very
large backgrounds and the $>$1 jet is also part of the low mass Higgs
searches.

Both CDF and D0 see no excesses with respect to the background
expectation in the statistical analysis of the respective MVA
discriminants and therefore set limits.  Using 8.2\,fb$^{-1}$ of data CDF
obtains an observed (expected) limit for $\mH$=165\,GeV of
\sigmaRatio=17.5(13.0).  For the muon +$\tau_{\rm h}$+$\etmiss$+$\leq$1 jet
and using 7.3\,fb$^{-1}$ D0 obtains a corresponding observed
(expected) limit of \sigmaRatio=6.6(7.8). The
lepton+$\tau$+$\etmiss$+$>1$ jet uses 4.3\,fb$^{-1}$ of D0 data and
obtains an observed (expected) limit for $\mH$=165\,GeV of
\sigmaRatio=12.4(12.3).

\subsection{Lepton+jets channel}
The signature of this channel is one isolated high $\pT$ electron or
muon, high $\etmiss$ and two high $\pT$ jets\,\cite{ref:lnjj}. For
$\mH$$\geq$ 160\,GeV, the Higgs decays to two on-shell $\rm W$ bosons,
thus offering the ability to reconstruct the kinematics of the full
event including the longitudinal momentum of the neutrino, up to a
two-fold ambiguity.  Although this signature suffers from large $\rm
W$+jets backgrounds, the large branching ratio of hadronic $\rm W$
boson decays provide 52 expected signal events surviving all
selections in 5.4\,fb$^{-1}$ of D0 data. The statistical analysis of
the MVA discriminant does not exhibit any excess with respect to the
background expectation and limits are set. Using 5.4\,fb$^{-1}$ of
data, D0 obtains an observed (expected) limit for $\mH$=165\,GeV of 
\sigmaRatio=5.2 (5.1).

\subsection{Same charge di-leptons and tri-leptons}\label{sec:ssDilepton}
In these channels, the defining characteristic is the presence of at
least two isolated, high $\pT$ electrons or muons, which can form a pair
of the same charge, and high $\etmiss$. Charge misidentification can
lead to a significant migration of opposite sign charge backgrounds
into the same sign region. Therefore high quality tracking criteria
are also required to suppress this instrumental background.  CDF also
requires the presence of at least 1 jet in the same charge di-lepton
final state, since the decay of the third boson will most likely
result in the production of an additional jet. 

\begin{figure*}[ht]
\begin{center}
\includegraphics[scale=0.38]{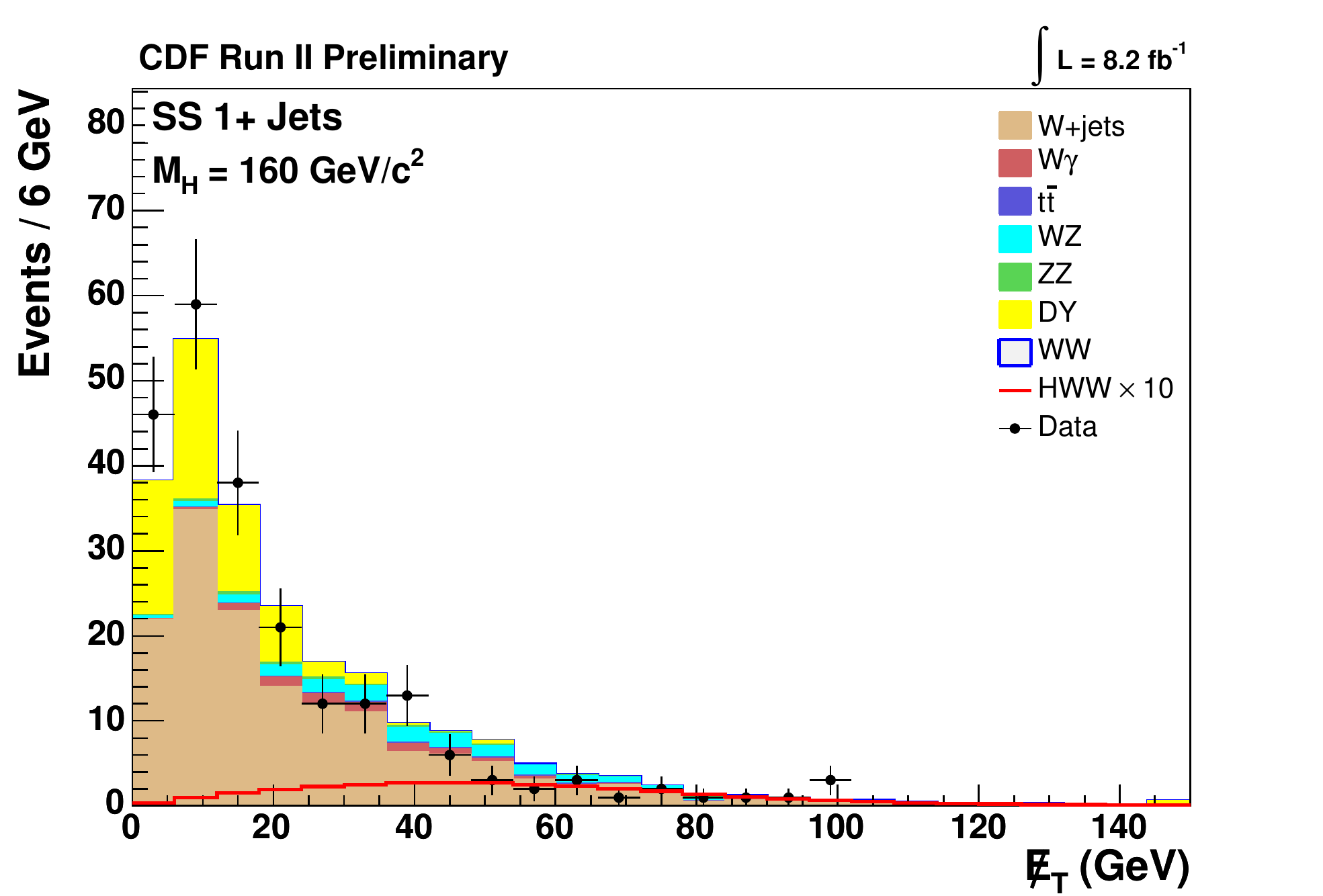}
\includegraphics[scale=0.38]{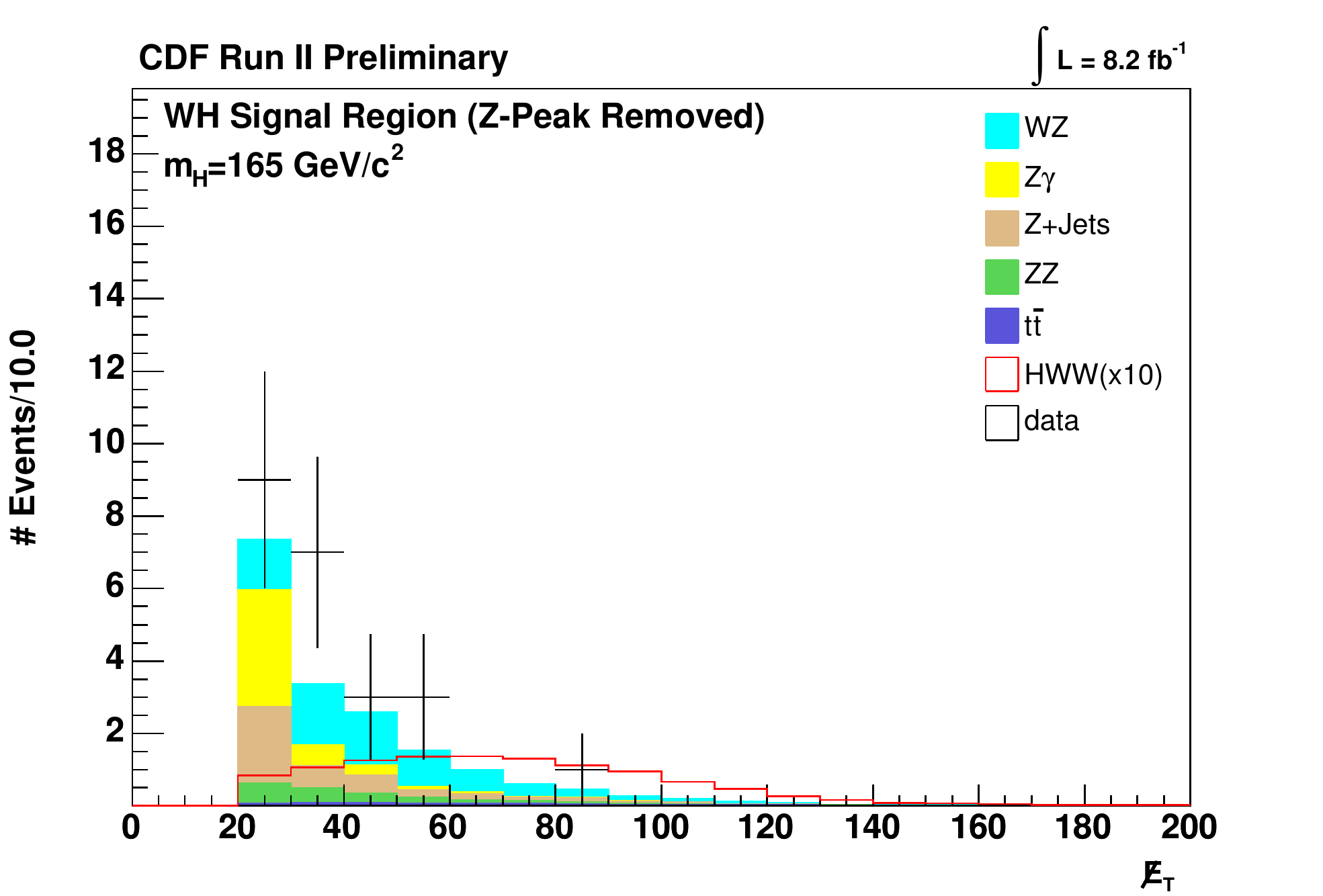}
\caption{$\etmiss$ distributions in same
  charge di-lepton events (left) tri-lepton events with no compatible Z mass (right).
\label{fig:ssDiTriLepton}}
\end{center}
\end{figure*}

In contrast to CDF, D0 does not include a dedicated tri-lepton search.
Tri-lepton events can occur naturally in $\mathrm{WH\to WWW}$ events
with all $\rm W$s decaying leptonically, or in $\mathrm{ZH\to ZWW}$
where the $\rm Z$ and one of the $\rm W$s decay leptonically and the
other hadronically. This distinction allows the separation of
tri-lepton events into cases where the same flavour opposite sign
leptons, form an invariant mass compatible to the $\rm Z$ mass or not.
If a mass compatible to the $\rm Z$ mass is found, the probability
that one of the $\rm W$s decayed hadronically, allows this channel to
be further split by requiring one or more than one reconstructed jets
in the event.  If more than one jets are found then the event is fully
reconstructed and the mass of the Higgs can be determined. CDF ensures
orthogonality between the same charge di-lepton and tri-lepton search
by vetoing the presence of a third lepton in the di-lepton case. As D0
has no dedicated tri-lepton search, no such veto is required.
Figure\,\ref{fig:ssDiTriLepton} shows the $\etmiss$ distributions in
same charge di-lepton and tri-lepton events after all selections.

No excess was observed in the statistical analysis of same charge
di-lepton and tri-lepton MVA discriminants, with respect to the
background expectation, and thus limits are set.  The same charge
di-lepton and tri-lepton limits of CDF are combined with the the
di-lepton+$\etmiss$ searches of Section\,\ref{sec:osDilepton}. The
most sensitive channel is the same charge di-lepton + $\etmiss$+$>1$
jets which gives an observed (expected) limit for $\mH$=165\,GeV of
\sigmaRatio=3.69(4.34) for 8.2\,fb$^{-1}$ of CDF data. The observed
(expected) limit of the same charge di-lepton channel of D0, using
5.4\,fb$^{-1}$ of data for $\mH$=165\,GeV is \sigmaRatio=6.4(7.3)\,\cite{ref:ss}.

\section{Systematic uncertainties}
The systematic uncertainties considered account for detector
resolutions, reconstruction efficiencies, background normalizations
and shapes, both theoretical and data-driven. Where appropriate, these
uncertainties are correlated between CDF and D0. Careful consideration
has also been taken to account for the theoretical uncertainties
related to the various Higgs production mechanisms and particularly
for $\mathrm{gg\to H}$, which is the main signal in the most sensitive
di-lepton+$\etmiss$ channel.

The Higgs boson signal is normalized to the most recent highest-order
calculations available for all production mechanisms considered. The
gluon fusion cross section, $\sigma(\mathrm{gg\to H})$, is calculated
to NNLO in QCD with soft gluon resummation to
NNLL\,\cite{ref:Anastasiou}\,\cite{ref:Grazzini} and uses the MSTW
2008 NNLO PDF set as it is the only NNLO set which results from a
global fit to all relevant data\,\cite{ref:MSTW}. Since the di-lepton
+$\etmiss$ analyses are split in categories depending on the number of
reconstructed jets, the QCD scale uncertainties on
$\sigma(\mathrm{gg\to H})$ are estimated following the prescreption
described in\,\cite{ref:errMatr}. By propagating the uncorrelated
uncertainties of the NNLL inclusive\,\cite{ref:Anastasiou}\,\cite{ref:Grazzini}, NLO one or more
jets\,\cite{ref:ggH01jetUncert} and NLO two or more
jets\,\cite{ref:ggH2jetUncert} rate to the exclusive $gg\to H+0$jet,
1jet and $>1$jet rate, the error matrix shown in
Table\,\ref{tab:ggHuncert} is built. The PDF uncertainty on
$\sigma(\mathrm{gg\to H})$, obtained using the prescription
in\,\cite{ref:ggH01jetUncert}\,\cite{ref:PDF4LHC}, 
is summarised in Table\,\ref{tab:ggHuncertPDF}.

\begin{table}
  \caption{Error matrix of the Scale ($\mu_{\rm r}$,$\mu_{\rm f}$) 
    uncertainties on $\sigma(\mathrm{gg\to H})$ for the three jet
    categories considered \label{tab:ggHuncert}.}
  \vspace{0.4cm}
  \begin{center}
  \begin{tabular}{cccc}
    \hline
    $\sigma$ $\mu_{R},\mu_{F}$ & $s_0$  & $s_1$ & $s_2$   \\
    \hline
    0jet & 13.4\% & -23.0\%  & -                        \\
    1jet & -      & 35.0\%   & -12.7\%                  \\
    $>$1jet & -      & -        & 33.0\%             \\
    \hline
  \end{tabular}
  \end{center}
\end{table}

\begin{table}[!]
  \caption{PDF uncertainties on $\sigma(\mathrm{gg\to H})$ 
    for the three jet categories considered \label{tab:ggHuncertPDF}.}
  \vspace{0.4cm}
  \begin{center}
    \begin{tabular}{cccc}
      \hline
      Uncertainty           & 0 jet & 1 jet  & $>$1 jet \\
      \hline
      PDF                   & 7.6\% & 13.8\% & 29.7\%   \\    
      \hline
    \end{tabular}
  \end{center}
\end{table}

\section{Results}
CDF and D0 set limits by combining all of the SM high mass channels
with up to 8.2\,fb$^{-1}$ of integrated luminosity. 
CDF excludes the range 156.5$<\mH<$173.7\,GeV (157$<\mH<$172.2\,GeV expected) 
and D0 excludes the range 161$<\mH<$170\,GeV (159$<\mH<$170\,GeV expected). 
In order to further improve the sensitivity of these searches, CDF and D0 combined their
individual high mass channels in the Tevatron high mass combination, which
is discussed elsewhere in these proceedings\,\cite{ref:TeVcombo}\,\cite{ref:TeVcurrent}.

{}

%Search for the standard model Higgs boson at LEP,
%\emph{Combined CDF and D0 Upper Limits on Standard Model Higgs-Boson Production with up to 8.1\,$fb^{-1}$ of Data}, 
%\emph{Precision Electroweak Measurements and Constraints on the Standard Model}, 
%\emph{Combined CDF and D0 Upper Limits on Standard Model Higgs-Boson Production with up to 8.6\,$fb^{-1}$ of Data}, 

%
%%\begin{thebibliography}
% and use \bibitem to create references.
%%\bibitem{RefJ}
% Format for Journal Reference
%%Author, Journal \textbf{Volume}, (year) page numbers
% Format for books
%%\bibitem{RefB}
%%Author, \textit{Book title} (Publisher, place year) page numbers
% etc
%%\end{thebibliography}

\end{document}